\documentclass[journal=jctcce,manuscript=article]{achemso}
\usepackage[utf8]{inputenc}
\usepackage{textcomp}
\usepackage{graphicx}
\usepackage[version=4]{mhchem}
\usepackage{siunitx}

\DeclareSIUnit\au{a.u.}
\usepackage{tabularx}
\DeclareSIUnit\angstrom{\text{\AA}}

\usepackage{comment}
\usepackage{csquotes}
\usepackage{amsmath}
\usepackage{tabularx}
\usepackage{import}
\usepackage{mathrsfs}  
\usepackage{multirow}
\usepackage[many]{tcolorbox}
\usepackage[flushleft]{threeparttable}
\usepackage{physics}
\allowdisplaybreaks[1]
\MakeOuterQuote{"}
\usepackage{mathtools,tikz,caption}
\DeclareRobustCommand\sampleline[1]{%
  \tikz\draw[#1] (0,0) (0,\the\dimexpr\fontdimen22\textfont2\relax)
  -- (1.5em,\the\dimexpr\fontdimen22\textfont2\relax);%
}
\newcommand{\group}[1]{\text{\textbf{\mbox{#1}}}}

\author{Tingting Zhao}
\author{Devin. A. Matthews}
\email{dmatthews@smu.edu}
\affiliation[Southern Methodist University]
{Department of Chemistry, Southern Methodist University, Dallas, TX}

\title{Analytic gradients for equation-of-motion coupled cluster with single, double, and perturbative triple excitations}

\definecolor{main}{HTML}{5989cf}    
\definecolor{sub}{HTML}{cde4ff}     
\definecolor{yellow}{HTML}{FFE4C4} 
\tcbset{
    sharp corners,
    colback = white,
    before skip = 0.2cm,    
    after skip = 0.5cm      
}    
\newtcolorbox{boxA}{
    colback = sub, 
    boxrule = 0pt  
}

\newtcolorbox{boxB}{
    colback = sub, 
    boxrule = 0pt  
}

\newtcolorbox{boxC}{
    colback = yellow, 
    boxrule = 0pt  
}

\newtcolorbox{boxD}{
    colback = yellow, 
    boxrule = 0pt  
}

\newtcolorbox{boxE}{
    colback = yellow, 
    boxrule = 0pt  
}

\begin{tocentry}
\includegraphics[scale=1]{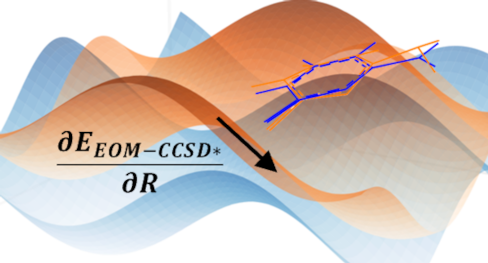}
\end{tocentry}

\begin{document}


\begin{abstract}
Understanding the process of molecular photoexcitation is crucial in various fields, including drug development, materials science, photovoltaics, and more. The electronic vertical excitation energy is a critical property, for example in determining the singlet-triplet gap of chromophores. However, a full understanding of excited-state processes requires additional explorations of the excited-state potential energy surface and electronic properties, which is greatly aided by the availability of analytic energy gradients. Owing to its robust high accuracy over a wide range of chemical problems, equation-of-motion coupled-cluster with single and double excitations (EOM-CCSD) is a powerful method for predicting excited state properties, and the implementation of analytic gradients of many EOM-CCSD (excitation energies, ionization potentials, electron attachment energies, etc.) along with numerous successful applications highlights the flexibility of the method. In specific cases where a higher level of accuracy is needed or in more complex electronic structures, the inclusion of triple excitations becomes essential, for example, in the EOM-CCSD* approach of Saeh and Stanton. In this work, we derive and implement for the first time the analytic gradients of EOMEE-CCSD*, which also provides a template for analytic gradients of related excited state methods with perturbative triple excitations. The capabilities of analytic EOMEE-CCSD* gradients are illustrated by several representative examples.
\end{abstract}

\section{Introduction}


The excited states of molecules exhibit a range of unique properties that are of great interest to many fields of science and technology: bioluminescence\cite{Gozem2017},  mutagenesis and carcinogenesis\cite{Chakraborty2018}, and photovoltaics and light-emitting diodes\cite{Kaloni2017, Zhang2019}, and many more. These diverse electronic characteristics arise from their complex potential energy landscapes, which can be probed through optical or ultraviolet (UV) absorption. The relaxation and decay processes to the ground state, occurring via either radiative or non-radiative pathways, further contribute to their functionality and provide valuable spectroscopic insights, such as fluorescence and phosphorescence. Theoretical quantum chemistry calculations play a crucial role in facilitating the interpretation of this experimental spectroscopy data, offering a comprehensive understanding of the excited states.

Despite the inherent complexity of excited states (for example, open-shell and multi-reference character), numerous models have been developed to calculate excited state energies and properties. 
Generally, these models fall into two categories. The first category comprises explicitly multi-reference methods that deal with solutions spanning a space of determinants or other suitable basis functions, which aim to provide either state-selective, state-averaged, or state-universal descriptions of electronic states. A comprehensive review of these methods is available elsewhere\cite{Lischka2018}. While these methods, such as complete active space self-consistent field (CASSCF) method\cite{Olsen2011}, its second perturbation-corrected variant (CASPT2)\cite{Andersson1990},  and the second-order $n$-electron valence state perturbation theory (NEVPT2)\cite{Angeli2001,Angeli2001-jcp,Angeli2002} have been highly successful, they demand a high level of expertise and familiarity with the system under investigation in order to tune the active orbital space and other calculation parameters. Additionally, interpreting the results can be a formidable challenge, as it often involves analyzing many interacting configurations and orbitals\cite{Bartlett2002,Lischka2018}. 
On the other hand, equation-of-motion coupled-cluster (EOM-CC) methods adopt a distinct approach, encompassing the entire set of multi-configurational target states within a single-reference framework. This black box theory has gained considerable popularity, owing to its favorable computational accessibility and its robust high accuracy over a wide range of chemical problems\cite{Watson2012}. For excited states, equation-of-motion coupled-cluster with single and double excitations (EOMEE-CCSD)\cite{Stanton1993} has become the gold standard method for the prediction of singly excited state energies and properties. The development of variants for ionization potentials (EOMIP-CCSD), electron attachment energies (EOMEA-CCSD), and spin-flip excitations (EOMSF-CCSD) has greatly extended its capability to describe doublet radicals, diradicals, triradicals, and bond breaking\cite{Krylov2005,musial_addition_2007}.

Energy differences and transition probabilities are the key quantities that are desirable to explain and predict the spectroscopic\cite{SerranoAndrs2005}. Importantly, these quantities have to be computed at significant points in the potential energy surface(PES). The development of analytic-derivative techniques has significantly facilitated the exploration of potential energy surfaces (PESs) of molecules in their electronic ground state, essentially for all available quantum-chemical schemes\cite{Gauss1991,Gauss1997,Gauss2002a,Hratchian2005,Kllay2003,Koch1990,Lee1991,Watts1992,Watts1993, Scuseria1991,Yamaguchi2011}. 
Gradients are accessible for various multireference methods, including MCSCF \cite{Busch1991}, MR-CI\cite{Page1984,Shepard1987}, and fully internally contracted CASPT2\cite{Vlaisavljevich2016}. Nevertheless, the efficiency of the EOM-CCSD gradient notably distinguishes itself. Analytic gradients for all the variants of EOM-CCSD methods have been documented\cite{Stanton1993,Stanton1994,Stanton_1994, Stanton1995,Levchenko2005,Pieniazek2008,Wladyslawski2005}.

While EOM-CCSD offers great accuracy for predominantly one-electron excitations, it is poor for double electron excitations\cite{Watts1994,Watts1995,Watts1996}. Efforts have been made in either iterative or non-iterative schemes to include triple excitation to improve accuracy, particularly for double-excited states \cite{Sattelmeyer2001}, EOM-CCSD*\cite{Stanton1996}, EOM-CCSD($\Tilde{\mathrm{T}}$)\cite{Watts1996}, EOM-CCSDT-3\cite{Watts1996}, EOM-CC3 (equivalent to LR-CC3 for energies)\cite{Christiansen1995,Koch1997}, and EOM-CCSDT\cite{Noga1987}. However, no analytic gradients have been published for these methods, although gradients for full EOMEE-CCSDT are available in the {\sc cfour} program package\cite{cfour} and transition dipole moments for EOM-CCSDT were reported by Hirata.\cite{10.1063/1.1753556} Counter-intuitively, the analytic gradients for the non-iterative triples methods are more complicated than for iterative models, as discussed below. In this paper, we present, for the first time, detailed formulas and implementation details of analytic closed-shell EOMEE-CCSD* energy gradients. Several demonstrative applications are also discussed.

\section{Theory}
 For convenience, in the following context $\ket{H}$ represents the set of all possible $n$-electron Slater determinants within a given spin-orbital basis. $\ket{H}$  can be further split into $\ket{H} =\ket{P} +\ket{Q} = \ket{0} +\ket{G} +\ket{Q} = \ket{0} +\ket{S} + \ket{D} +\ket{Q}$, where $\ket{0}$ is the reference determinant (Fermi vacuum), $\ket{S} =\left|_{i}^{a}\right\rangle$ are the singly-excited determinants, and $\ket{D}=\left|_{ij}^{ab}\right\rangle$ are the doubly-excited determinants.  $\ket{Q} =\left|_{ijk}^{abc}\right\rangle +\left|_{ijkl}^{abcd}\right\rangle +....$,  represents any determinant of excitation rank higher than two, out of which we specifically identify the triply-excited determinants $\ket{T}=\left|_{ijk}^{abc}\right\rangle$.

Briefly, the ground CCSD and excited state EOMEE-CCSD energies are given by,
\begin{align}
E_{\text{CCSD}} =& \bra{0}e^{-\hat{T}}\hat{H}e^{\hat{T}}\ket{0} = \bra{0}(\hat{H}e^{\hat{T}})_c\ket{0} = \bra{0}\bar{H}\ket{0} \label{eq:cc_energy} \\
0 =& \bra{G}\bar{H}\ket{0} \label{eq:t_amp} \\
\hat{H} =& \sum_{pq} f^p_q \{ a^\dagger_p a_q \} + \frac{1}{4}\sum_{pqrs} v^{pq}_{rs} \{ a^\dagger_p a^\dagger_q a_s a_r \} = \hat{F} + \hat{V} \\
\hat{T} =& \sum_{ai} t^a_i a^\dagger_a a_i + \frac{1}{4}\sum_{abij} t^{ab}_{ij} a^\dagger_a a^\dagger_b a_j a_i = \hat{T}_1 + \hat{T}_2 \\
E_{\text{EOM-CCSD}} =& \bra{0}\hat{L}\bar{H}\hat{R}\ket{0} \label{eq:eom_energy} \\
0 =& \bra{G}(\bar{H}-E_{\text{EOM-CCSD}})\hat{R}\ket{0} \label{eq:r_amp} \\
0 =& \bra{0}\hat{L}(\bar{H}-E_{\text{EOM-CCSD}})\ket{G} \label{eq:l_amp} \\
\hat{R} =& r_0 + \sum_{ai} r^a_i a^\dagger_a a_i + \frac{1}{4}\sum_{abij} r^{ab}_{ij} a^\dagger_a a^\dagger_b a_j a_i = \hat{R}_0 + \hat{R}_1 + \hat{R}_2 \\
\hat{L} =& \sum_{ai} l^i_a a^\dagger_i a_a + \frac{1}{4}\sum_{abij} l^{ij}_{ab} a^\dagger_i a^\dagger_j a_b a_a = \hat{L}_1 + \hat{L}_2
\end{align}
where $\{\ldots\}$ indicates normal ordering and $(\ldots)_c$ indicates a connected operator product. The energy $E_{\text{EOM-CCSD}}$ and amplitudes $\hat{R}$ and $\hat{L}$ correspond to the same excited state which is otherwise unspecified, except that it should be non-degenerate. When properties of a specific excited state $\mu$ are required, we will employ the notation $E_{\text{EOM-CCSD}}(\mu)$, $\hat{R}(\mu)$, $\hat{L}(\mu)$, etc.

The perturbative third-order correction\cite{Saeh1999} to the EOMEE-CCSD energy can be written as follows, with $\omega=E_{\text{EOM-CCSD}}-E_{\text{CCSD}}$ being the EOMEE-CCSD vertical excitation energy,
\begin{align}
\Delta E =&\bra{0}\hat{L}\hat{V}\ket{T} \bra{T}(\omega - \hat{F})^{-1}\ket{T} \bra{T}\hat{V}\hat{R}_{2}+(\hat{V} \hat{T}_{2}\hat{R}_{1})_c\ket{0} 
\end{align}
The total EOM-CCSD* energy can be written as,
\begin{align}
E_{\text{EOM-CCSD*}} =&E_{\text{EOM-CCSD}}+\Delta E=\bra{0}\hat{L}\bar{H}\hat{R}\ket{0} +\bra{0}\hat{L}_{3}(\omega - \hat{F})\hat{R}_{3}\ket{0}  \\
\bra{0}\hat{L}_{3}\ket{T} =&\bra{0}\hat{L}\hat{V}(\omega - \hat{F})^{-1}\ket{T} \\
\bra{T}\hat{R}_{3}\ket{0} =&\bra{T}(\omega - \hat{F})^{-1}(\hat{V}\hat{R}_{2}+(\hat{V}\hat{T}_{2}\hat{R}_{1})_c)\ket{0}
\end{align}
assuming a canonical reference determinant. Straightforward differentiation with respect to an arbitrary perturbation $\chi$ gives, 
\begin{align} \label{eq:all_0}
E_{\text{EOM-CCSD*}}^{\chi} =& E_{\text{EOM-CCSD}}^{\chi}
+\bra{0}\hat{L}^{\chi}\hat{V}\hat{R}_{3}\ket{0} +\bra{0}\hat{L}\hat{V}^{\chi}\hat{R}_{3}\ket{0} \nonumber \\
&- \bra{0}\hat{L}_{3}(\omega - \hat{F})^{\chi}\hat{R}_{3}\ket{0} + \bra{0}\hat{L}_{3}(\hat{V}\hat{R}_{2}^{\chi}+\hat{V}\hat{T}_{2}\hat{R}_{1}^{\chi})\ket{0} \nonumber \\
&+ \bra{0}\hat{L}_3\hat{V}\hat{R}_{1}\hat{T}_{2}^{\chi}\ket{0} + \bra{0}\hat{L}_{3}\hat{V}^{\chi}(\hat{R}_{2}+\hat{T}_{2}\hat{R}_{1})\ket{0}  
\end{align}
Note there is a negative sign in the term $\bra{0}\hat{L}_{3}(\omega - \hat{F})^{\chi}\hat{R}_{3}\ket{0}$ as,
\begin{align}
((\omega - \hat{F})^{-1})^{\chi}= - (\omega - \hat{F})^{-1} (\omega - \hat{F})^{\chi}(\omega - \hat{F})^{-1} \label{eu_eqn}
\end{align}
This term can be further expanded as follows, where for convenience we set $\delta =\bra{0}\hat{L}_{3}\hat{R_{3}}\ket{0}$,
\begin{align}
-\bra{0}\hat{L}_{3}(\omega-\hat{F})^{\chi}\hat{R}_{3}\ket{0}=&\bra{0}\hat{L}_{3}(\hat{F}^{\chi}+E_{\text{CCSD}}^{\chi}-E_{\text{EOM-CCSD}}^{\chi})\hat{R}_{3}\ket{0} \nonumber \\
=&\bra{0}\hat{L}_{3}\hat{F}^{\chi}\hat{R}_{3}\ket{0} + \delta (E_{\text{CCSD}}^{\chi}- E_{\text{EOM-CCSD}}^{\chi} )\label{denor_term}
\end{align}

$E_{\text{CCSD}}^{\chi}$  and $E_{\text{EOM-CCSD}}^{\chi}$ can be obtained by directly differentiating the energy equations \eqref{eq:cc_energy} and \eqref{eq:eom_energy}, respectively. Noting the amplitude equations \eqref{eq:t_amp}, \eqref{eq:r_amp}, and \eqref{eq:l_amp} and that $(\bra{0}\hat{L}\hat{R}\ket{0})^\chi = 0$,
\begin{align}
 E_{\text{CCSD}}^{\chi} =&\bra{0}\bar{H}\hat{T}^{\chi}\ket{0} +\bra{0}\bar{H}^{(\chi)}\ket{0}  \label{equation:ccsd_chi} \\
E_{\text{EOM-CCSD}}^{\chi}=&\bra{0}\hat{L}\bar{H}^{(\chi)}\hat{R}\ket{0}+\bra{0}\hat{L}\bar{H}\ket{Q}\bra{Q}\hat{R}\hat{T}^{\chi}\ket{0} 
\label{EOMderivative}
\end{align}
where $\bar{H}^{(\chi)} = (\hat{H}^\chi e^{\hat{T}})_c$.

Combining \eqref{eq:all_0}, \eqref{denor_term}, \eqref{equation:ccsd_chi}, and \eqref{EOMderivative} and grouping all terms into four parts we obtain,
\begin{align} 
E_{\text{EOM-CCSD*}}^{\chi} =& \group{I} +\group{II}+\group{III}+\group{IV} \nonumber \\
\group{I} =& \bra{0}\hat{L}^{\chi}\hat{V}\hat{R}_{3}\ket{0} \\
\group{II} =& \bra{0}\hat{L}_{3} (\hat{V}\hat{R}_{2}^{\chi}+\hat{V}\hat{T}_{2}\hat{R}_{1}^{\chi})\ket{0} \\
\group{III} =& \bra{0}\hat{L}_{3}(\hat{V}\hat{T}_{2}^{\chi}\hat{R}_{1})\ket{0} +\bra{0}(1-\delta)\hat{L}\bar{H}\ket{Q}\bra{Q}\hat{R}\hat{T}^{\chi}\ket{0}+\delta\bra{0}\bar{H}\hat{T}^{\chi}\ket{0} \\
\group{IV} =& \bra{0}(1-\delta)\hat{L}\bar{H}^{(\chi)}\hat{R}\ket{0} 
+\bra{0}\hat{L}_{3}(\hat{V}^{\chi}\hat{R}_{2}+\hat{V}^{\chi}\hat{T}_{2}\hat{R}_{1})\ket{0} \nonumber \\
&+\bra{0}\hat{L}\hat{V}^{\chi}\hat{R}_{3}\ket{0} 
+\bra{0}\hat{L}_3 \hat{F}^{\chi} \hat{R}_3\ket{0} +\delta\bra{0}\bar{H}^{(\chi)}\ket{0} 
\end{align}
The terms are grouped such that \group{I} depends (directly) only on $\hat{L}^{\chi}$, \group{II} depends only on $\hat{R}^{\chi}$, \group{III} depends only on $\hat{T}^{\chi}$, and \group{IV} depends only on the derivatives of integrals.

$\hat{L}^\chi$ and $\hat{R}^\chi$ of course depend in turn on $\hat{T}^{\chi}$ due to the presence of $\bar{H}$ in their definite amplitude equations. Thus, we should tackle \group{I} and \group{II} before moving on to \group{III}. To analyze \group{I}, we first define a convenient intermediate, noting that $\hat{L}^\chi$ spans only the $\bra{G}$ space,
\begin{align}
\group{I}=&\bra{0}\hat{L}^{\chi}\hat{V}\hat{R}_{3}\ket{0} =\bra{0}\hat{L}^{\chi}\hat{\Sigma}\ket{0} \\
\bra{G}\hat{\Sigma}\ket{0} =& \bra{G}\hat{V}\hat{R}_{3}\ket{0}
\end{align}
We then introduce a definition of $\hat{L}^\chi$ by differentiating \eqref{eq:l_amp},
 \begin{align}
\bra{0}\hat{L}^{\chi}(\bar{H}-E_{\text{EOM-CCSD}})\ket{G} =& \bra{0}\hat{L}\hat{T}^{\chi}(\bar{H}-E_{\text{EOM-CCSD}})\ket{G}- \bra{0}\hat{L}\bar{H}^{(\chi)}\ket{G} \nonumber \\
&+ E_{\text{EOM-CCSD}}^{\chi} \bra{0}\hat{L}\ket{G} - \bra{0}\hat{L}\bar{H}\ket{Q} \bra{Q}\hat{T}^{\chi}\ket{G}
\label{Lderivative}
\end{align}

Notionally, one would utilize this definition by applying the inverse of $\bra{G}\bar{H}-E_{\text{EOM-CCSD}}\ket{G}$ to both sides of the equation, yielding $\bra{0}\hat{L}^{\chi}\ket{G} = \bra{0}\hat{X}\ket{G}\bra{G}\bar{H}-E_{\text{EOM-CCSD}}\ket{G}^{-1}$ with $\hat{X}$ collecting all terms from the R.H.S., which then could be substituted in \group{I} giving $\bra{0}\hat{X}\ket{G}\bra{G}\bar{H}-E_{\text{EOM-CCSD}}\ket{G}^{-1}\bra{G}\hat{\Sigma}\ket{0}$. Finally, one would then define a new set of amplitudes $\bra{G}\hat{\Upsilon}\ket{0}$ defined by the system of equations $\bra{G}(\bar{H}-E_{\text{EOM-CCSD}})\hat{\Upsilon}\ket{0} = \bra{G}\hat{\Sigma}\ket{0}$.
However, this plan is hindered by the fact that $E_{\text{EOM-CCSD}}$ is an exact eigenvalue of $\bra{G}\bar{H}\ket{G}$, and thus the shifted matrix is exactly singular. This problem can be avoided by imposing an additional condition on $\hat{L}^\chi$ (and $\hat{R}^\chi$): while the normalization condition $\bra{0}\hat{L}\hat{R}\ket{0}=1$ ensures that $\bra{0}\hat{L}^\chi\hat{R}\ket{0} + \bra{0}\hat{L}\hat{R}^\chi\ket{0} = 0$, we can additionally require a biorthogonal form of intermediate normalization such that $\bra{0}\hat{L}^\chi\hat{R}\ket{0} = \bra{0}\hat{L}\hat{R}^\chi\ket{0} = 0$. Thus, the part of $\hat{\Sigma}$ parallel to $\hat{R}$ is not required and can be projected out. Similarly, the fact that $\hat{L}$ lies in the null space of $\bra{G}(\bar{H}-E_{\text{EOM-CCSD}})\ket{G}$ implies that each term on the R.H.S. of \eqref{Lderivative} is perpendicular to $\hat{L}$. Thus, it is sufficient and safe to employ the pseudo-inverse,
\begin{align}
\bra{0}\hat{L}^{\chi}\ket{G} =& \bra{0}\hat{L}\hat{T}^{\chi}\ket{G}(\hat{I}-\bra{G}\hat{R}\ket{0} \bra{0}\hat{L}\ket{G}) \nonumber \\
&- \left[ \bra{0}\hat{L}\bar{H}^{(\chi)}\ket{G} - E_{\text{EOM-CCSD}}^{\chi} \bra{0}\hat{L}\ket{G} \right. \nonumber \\
&\left. + \bra{0}\hat{L}\bar{H}\ket{Q} \bra{Q}\hat{T}^{\chi}\ket{G} \right]  \bra{G}\bar{H}-E_{\text{EOM-CCSD}}\ket{G}^+ \label{Lderivative2} \\
\bra{G}\bar{H}-E_{\text{EOM-CCSD}}(\mu)\ket{G}^+ =& \sum_{\nu \ne \mu} \frac{\bra{G}\hat{R}(\nu)\ket{0}\bra{0}\hat{L}(\nu)\ket{G}}{E_{\text{EOM-CCSD}}(\nu) - E_{\text{EOM-CCSD}}(\mu)}
\end{align}
where the pseudo-inverse satisfies,
\begin{align}
\bra{G}\bar{H}-E_{\text{EOM-CCSD}}\ket{G} \bra{G}\bar{H}-E_{\text{EOM-CCSD}}\ket{G}^+ =&  \nonumber \\
\bra{G}\bar{H}-E_{\text{EOM-CCSD}}\ket{G}^+ \bra{G}\bar{H}-E_{\text{EOM-CCSD}}\ket{G} =& \, \hat{I} - \bra{G}\hat{R}\ket{0} \bra{0}\hat{L}\ket{G}
\end{align}
with $\hat{I}$ being the identity operator.

The introduction of the pseudo-inverse now allows $\hat{L}^\chi$ to be substituted into \group{I},
\begin{align}
\group{I} =& \bra{0}\hat{L}\hat{T}^{\chi}\hat{\Sigma}^\perp\ket{0} - \left[ \bra{0}\hat{L}\bar{H}^{(\chi)}\ket{G} - E_{\text{EOM-CCSD}}^{\chi} \bra{0}\hat{L}\ket{G} \right. \nonumber \\
&\left. + \bra{0}\hat{L}\bar{H}\ket{Q} \bra{Q}\hat{T}^{\chi}\ket{G} \right]  \bra{G}\bar{H}-E_{\text{EOM-CCSD}}\ket{G}^+ \bra{G}\hat{\Sigma}\ket{0} \nonumber \\
=& \bra{0}\hat{L}\hat{\Sigma}^\perp\hat{T}^{\chi}\ket{0} + \bra{0}\hat{L}\bar{H}^{(\chi)}\hat{\Upsilon}\ket{0} - E_{\text{EOM-CCSD}}^\chi \bra{0}\hat{L}\hat{\Upsilon}\ket{0} \nonumber \\
& + \bra{0}\hat{L}\bar{H}\ket{Q}\bra{Q}\hat{\Upsilon}\hat{T}^{\chi}\ket{0} \\
\bra{G}\hat{\Sigma}^\perp\ket{0} =& \bra{G}\hat{\Sigma}\ket{0} - \bra{G}\hat{R}\ket{0} \bra{0}\hat{L}\hat{\Sigma}\ket{0}
\end{align}
where we introduce an excitation operator $\hat{\Upsilon}$ spanning $\ket{G}$ as,
\begin{align}
\bra{G}(\bar{H}-E_{\text{EOM-CCSD}}) \hat{\Upsilon}\ket{0} =& -\bra{G}\hat{\Sigma}^\perp\ket{0}
\end{align}
The portion of $\hat{\Upsilon}$ parallel to $\hat{R}$ is undefined, but it is convenient and numerically stable to require $\hat{\Upsilon}$ be perpendicular to $\hat{R}$. In particular, this means that $\bra{0}\hat{L}\hat{\Upsilon}\ket{0} = 0$. Finally we note that biorthogonal intermediate normalization is only one choice of uniquely determining $\bra{0}\hat{L}^\chi\hat{R}\ket{0}$ and $\bra{0}\hat{L}\hat{R}^\chi\ket{0}$. For example, one could require that $\bra{0}\hat{R}^\dagger\hat{R}\ket{0}=1$ which would still allow the definition of a consistent set of $\hat{\Upsilon}$ amplitudes with a modified set of equations (involving an operator more complicated than the pseudo-inverse). Thus to avoid a loss of generality we do not assume that $\bra{0}\hat{L}\hat{\Upsilon}\ket{0} = 0$ or other similar consequences of biorthogonal intermediate normalization.

For $\group{II}$, we employ a similar strategy,
\begin{align}
\group{II}=&\bra{0}\hat{L}_{3}(\hat{V}\hat{R}_{2}^{\chi}+\hat{V}\hat{T}_{2}\hat{R}_{1}^{\chi})\ket{0} = \bra{0}\hat{\Omega}\hat{R}^{\chi}\ket{0} \\
\bra{0}\hat{\Omega}\ket{S}=&\bra{0}\hat{L}_{3}\hat{V}\hat{T}_{2}\ket{S} \\
\bra{0}\hat{\Omega}\ket{D}=&\bra{0}\hat{L}_{3}\hat{V}\ket{D} \\
 \bra{G}\hat{R}^{\chi}\ket{0} =& -(\hat{I}-\bra{G}\hat{R}\ket{0} \bra{0}\hat{L}\ket{G})\bra{G}\hat{R}\hat{T}^{\chi}\ket{0} \nonumber \\
 &- \bra{G}\bar{H}-E_{\text{EOM-CCSD}}\ket{G}^+ \left[ \bra{G}\bar{H}^{(\chi)}\hat{R}\ket{0} \right. \nonumber \\
 & \left. - E_{\text{EOM-CCSD}}^{\chi}\bra{G}\hat{R}\ket{0} + \bra{G}\bar{H}\ket{Q}\bra{Q}\hat{R}\hat{T}^\chi\ket{0} \right]    \label{Rderivative} 
\end{align}
Substitution followed by the definition of another set of de-excitation amplitudes $\hat{\Pi}$ gives,
\begin{align}
\group{II}=&-\bra{0}\hat{\Omega}^\perp\hat{R}\hat{T}^{\chi}\ket{0} -\bra{0}\hat{\Omega}\ket{G} \bra{G}\bar{H}-E_{\text{EOM-CCSD}}\ket{G} ^+  \nonumber \\
& \times \left[ \bra{G}\bar{H}^{(\chi)}\hat{R}\ket{0} - E_{EOM}^{\chi}\bra{G}\hat{R}\ket{0} + \bra{G}\bar{H}\ket{Q}\bra{Q}\hat{R}\hat{T}^{\chi}\ket{0} \right] \nonumber \\
=&-\bra{0}\hat{\Omega}^\perp\hat{R}\hat{T}^{\chi}\ket{0} +\bra{0}\hat{\Pi}\bar{H}^{(\chi)}\hat{R}\ket{0} -E_{\text{EOM-CCSD}}^{\chi}\bra{0}\hat{\Pi}\hat{R}\ket{0}
\nonumber \\
&   + \bra{0}\hat{\Pi}\bar{H}\ket{Q}\bra{Q}\hat{R}\hat{T}^\chi\ket{0}
\end{align}
with $\hat{\Pi}$ defined as,
\begin{align}
\bra{0}\hat{\Pi}(\bar{H}-E_{\text{EOM-CCSD}})\ket{G} =&-\bra{0}\hat{\Omega}^\perp\ket{G}
\end{align}

Combining \group{I} and \group{II} and expanding $E_{\text{EOM-CCSD}}^{\chi}$ using \eqref{EOMderivative}, we arrive at an expression where the direct dependence on $\hat{R}^\chi$ and $\hat{L}^\chi$ has been replaced by dependence on $\hat{T}^\chi$ and $\hat{H}^\chi$ (via $\bar{H}^{(\chi)}$),
\begin{align}
\group{I}+\group{II}=& \left[ \bra{0}(\hat{L}\hat{\Sigma}^\perp-\hat{\Omega}^\perp\hat{R})\hat{T}^{\chi}\ket{0}
+\bra{0}\hat{L}\bar{H}\ket{Q} \bra{Q}\hat{\Upsilon} \hat{T}^{\chi}\ket{0} \right.
 \nonumber \\ 
& \left. +\bra{0}\hat{\Pi}\bar{H}\ket{Q} \bra{Q}\hat{R}\hat{T}^{\chi}\ket{0}
-\varepsilon \bra{0}\hat{L}\bar{H}\ket{Q} \bra{Q}\hat{R}\hat{T}^{\chi}\ket{0} \right] \nonumber \\
&+\left[ \bra{0}\hat{L}\bar{H}^{(\chi)}\hat{\Upsilon}\ket{0} +\bra{0}\hat{\Pi}\bar{H}^{(\chi)}\hat{R}\ket{0}
-\varepsilon \bra{0}\hat{L}\bar{H}^{(\chi)}\hat{R}\ket{0} \right] \nonumber \\
=& \,\group{III${}^\prime$} + \group{IV${}^\prime$} \\
\varepsilon=&\bra{0}(\hat{\Pi}\hat{R}+\hat{L}\hat{\Upsilon})\ket{0}
\end{align}
Note that based on the assumptions above, $\varepsilon = 0$. However, we keep $\varepsilon$ in the following derivation for completeness.

The additional terms in \group{III${}^\prime$} can now be added to \group{III},
\begin{align}
  \group{III} + \group{III${}^\prime$} =& \bra{0}\hat{L}_{3}\hat{V}\hat{R}_{1}\hat{T}_{2}^{\chi}\ket{0} +\bra{0}(1-\delta-\varepsilon)\hat{L}\bar{H}\ket{Q}\bra{Q}\hat{R}\hat{T}^{\chi}\ket{0}+\delta\bra{0}\bar{H}\hat{T}^{\chi}\ket{0} \nonumber \\
  &+\bra{0}(\hat{L}\hat{\Sigma}^\perp-\hat{\Omega}^\perp\hat{R})\hat{T}^{\chi}\ket{0}
+\bra{0}\hat{L}\bar{H}\ket{Q} \bra{Q}\hat{\Upsilon} \hat{T}^{\chi}\ket{0} 
 \nonumber \\ 
&  +\bra{0}\hat{\Pi}\bar{H}\ket{Q} \bra{Q}\hat{R}\hat{T}^{\chi}\ket{0} \nonumber \\
=& \bra{0}\hat{\Xi}\hat{T}^\chi\ket{0} \\
\bra{0}\hat{\Xi}\ket{G} =& \bra{0}\hat{L}_{3}\hat{V}\hat{R}_{1}\ket{D} + \bra{0}(1-\delta-\varepsilon)\hat{L}\bar{H}\ket{Q}\bra{Q}\hat{R}\ket{G}+\delta\bra{0}\bar{H}\ket{G} \nonumber \\
  &+\bra{0}(\hat{L}\hat{\Sigma}^\perp-\hat{\Omega}^\perp\hat{R})\ket{G}
+\bra{0}\hat{L}\bar{H}\ket{Q} \bra{Q}\hat{\Upsilon}\ket{G} 
 \nonumber \\ 
&  +\bra{0}\hat{\Pi}\bar{H}\ket{Q} \bra{Q}\hat{R}\ket{G}
\end{align}
Differentiation of the $\hat{T}$ amplitude equations \eqref{eq:t_amp} yields an expression for $\bra{G}\hat{T}^{\chi}\ket{0}$, where a pseudo-inverse is no longer required due to the fact that while $E_{\text{CCSD}} = \bra{0}\bar{H}\ket{0}$ is an eigenvalue of $\bra{P}\bar{H}\ket{P}$, removal of the reference determinant in $\bra{G}\bar{H}-E_{\text{CCSD}}\ket{G}$ prevents singularity,
\begin{align}
\bra{G}\hat{T}^{\chi}\ket{0} = -\bra{G}\bar{H} -E_{\text{CCSD}}\ket{G}^{-1} \bra{G}\bar{H}^{(\chi)}\ket{0}
\end{align}
Using this definition we introduce a final set of de-excitation amplitudes $\hat{\mathcal{Z}}$,
\begin{align}
\group{III} + \group{III${}^\prime$} =& - \bra{0}\hat{\Xi}\ket{G}\bra{G}\bar{H} -E_{\text{CCSD}}\ket{G}^{-1} \bra{G}\bar{H}^{(\chi)}\ket{0} \nonumber \\
=& \bra{0}\hat{\mathcal{Z}} \bar{H}^{(\chi)}\ket{0} \nonumber \\
=& \,\group{IV${}^{\prime\prime}$} \\
\bra{0}\hat{\mathcal{Z}}(\bar{H}-E_{\text{CCSD}})\ket{G} =& -\bra{0}\hat{\Xi}\ket{G}
\end{align}

Finally, the remaining terms \group{IV${}^{\prime}$} and \group{IV${}^{\prime\prime}$} from above can be combined with \group{IV} to yield the total gradient expression,
\begin{align}
E_{\text{EOM-CCSD*}}^{\chi}=&\group{IV} + \group{IV${}^{\prime}$} + \group{IV${}^{\prime\prime}$}\nonumber \\
=& \bra{0}\hat{\Pi}\bar{H}^{(\chi)}\hat{R}\ket{0} +\bra{0}\hat{L}\bar{H}^{(\chi)}\hat{\Upsilon}\ket{0} \nonumber \\
&+(1-\delta-\varepsilon)\bra{0}\hat{L}\bar{H}^{(\chi)}\hat{R}\ket{0} +\bra{0}(\delta + \hat{\mathcal{Z}})\bar{H}^{(\chi)}\ket{0} \nonumber \\
&+\bra{0}\hat{L}\hat{V}^{\chi}\hat{R}_{3}\ket{0} + \bra{0}\hat{L}_{3}(\hat{V}^{\chi}\hat{R}_{2}+\hat{V}^{\chi}\hat{T}_{2}\hat{R}_{1})\ket{0} \nonumber \\
&+\bra{0}\hat{L}_3\hat{F}^{\chi}\hat{R}_3\ket{0} \\
=& \sum_{p} D^p_p (\epsilon_p)^\chi + \sum_{pqrs} \Gamma^{pq}_{rs} (v^{pq}_{rs})^\chi
\end{align}
where $\epsilon_p = f^p_p$. The one- and two-particle density matrices $D^p_q$ and $\Gamma^{pq}_{rs}$ can be then be constructed from the various amplitudes and then processed using standard techniques such as in the computation of the CCSD(T) gradient\cite{Scuseria1991,Hald2003}.

\section{Results}

The analytic gradients of EOMEE-CCSD* have been implemented for closed-shell reference determinants in the development version of the {\sc cfour} program package.\cite{cfour} In this section, we discuss the validation of our implementation and demonstrate the illustrative application of analytic EOMEE-CCSD* gradients to a set of excited states.
In the geometry optimization, the following convergence thresholds were used, denoted using the relevant {\sc cfour} keywords: \texttt{SCF\_CONV} = $10^{-10}$, \texttt{CC\_CONV} = $10^{-9}$, \texttt{LINEQ\_CONV}  = $10^{-8}$, \texttt{ESTATE\_CONV} = $10^{-8}$, \texttt{GEO\_MAX\_STEP} = 50 a.u., and \texttt{GEO\_CONV} = $10^{-7}$, except where indicated.

\subsection{Validation}

The excited states of formaldehyde have been extensively investigated both experimentally\cite{Moule1975} and theoretically\cite{Merchn1995,Hachey1996,Bruna1997,DelBene1998,Dallos2001,LISCHKA2002}. In addition, its small system size makes it a great candidate for validating the correctness of our implementation against numerical differentiation. The $1{}^1\text{A}_2(n\xrightarrow{}\pi^*)$ valence excited state is investigated here. The ground state geometry is obtained from the QUEST2 dataset\cite{Loos2019}, which has been optimized at the CC3/aug-cc-pVTZ level. The computed EOM-CCSD* gradients (with the aug-cc-pVDZ basis) using both the analytic gradient we implemented and finite differences of energies agree well with each other. The differences between the numerical and analytic values for the individual gradient components were, in all cases, less than $10^{-6}$ a.u. (see Table 1 in the Supporting Information), which confirmed the correctness of our implementation.

It has been established that the $1{}^1\text{A}_2$ state exhibits a non-planar equilibrium geometry\cite{Moule1975,Merchn1995,LISCHKA2002}. Consequently, $\text{C}_\text{s}$ symmetry is employed during optimization. The optimized geometries and adiabatic excitation energies are summarized in Table~\ref{tab:formaldehyde_geometry}. The optimized geometries are compared to lower-level EOM-CCSD and higher-level EOM-CCSDT. To facilitate comparison, experimental data are also included\cite{Job1969,Jensen1982,Lessard1977,Clouthier1983}. 
As expected, optimized geometries with EOM-CCSDT are in better agreement with the experimental data. One finds $\leq 0.01$ \AA{} difference in the \ce{C-O} bond and \ce{C-H} bonds, $\ang{0.4}$ and $\ang{2.1}$ difference in the angle and out-of-plane dihedral, respectively.

\begin{table}[hb]
    \caption{Optimized geometry of $n\xrightarrow{}\pi^*$ excited state of formaldehyde with cc-aug-pVTZ basis set.}
    \label{tab:formaldehyde_geometry}
    \centering
\begin{threeparttable}
    \begin{tabular}{cSSSS[table-align-text-post = false]}
\hline
\hline
& \multicolumn{1}{c}{EOM-CCSD} & \multicolumn{1}{c}{EOM-CCSD*} & \multicolumn{1}{c}{EOM-CCSDT} & \multicolumn{1}{c}{Expt.} \\
\hline
$r_\text{CO}$ (\AA)&1.306&1.343&1.331&1.323${}^{a}$\\
$r_\text{CH}$ (\AA)&1.091&1.089&1.093&1.103${}^{a}$\\
$\angle_\text{COH}$ (${}^\circ$) &119.0&121.0&118.5&118.1${}^a$\\
Out of plane dihedral (${}^\circ$)&30.4&27.8&36.1&34.0${}^{a}$\\
$T_e$ (eV) & 3.72  & 3.12 & 3.06 &3.50${}^{b}$ \\
\hline
\hline
    \end{tabular}
\begin{tablenotes}
    \item{a)} Ref.~\citenum{Jensen1982}.
    \item{b)} Ref.~\citenum{Lessard1977}.
\end{tablenotes}
\end{threeparttable}
\end{table}

\begin{table}[ht]
\caption{Harmonic vibrational frequencies ($\omega$), infrared intensities ($I$) for the  $n\xrightarrow{}\pi^*$ excited state of formaldehyde with cc-aug-pVDZ. Numbering of normal modes follows the assignment in the ground state. } \label{tab:formaldehyde-frequency}
\centering
\begin{threeparttable}
    \begin{tabular}{cSSSSSSS[table-align-text-post = false]}
    \hline
    \hline
    \multicolumn{1}{l}{ }    & \multicolumn{2}{c}{EOM-CCSD}  & \multicolumn{2}{c}{EOM-CCSD*}  & \multicolumn{2}{c}{EOM-CCSDT} & \multicolumn{1}{c}{Expt.${}^{a}$}  \\
    Mode & $\omega$ &  $I$ & $\omega$ &  $I$ &$\omega$ &  $I$ &$\omega$\\ \hline
$\omega_1 (A^{\prime})$   & 3143.1 & 6.4 & 3054.4 & 0.5 & 3011.5 & 3.0&2846\\  
$\omega_2 (A^{\prime})$  & 1291.7 & 30.0 & 1131.7 & 68.9 & 1198.1 & 21.7&1183\\ 
$\omega_3 (A^{\prime})$   & 1362.0 & 5.1 & 1364.7 & 0.7 & 1342.8 & 8.2&1293.1\\
$\omega_4 (A^{\prime})$   & 427.0 &  52.2 & 548.7 & 38.1& 655.8 & 40.2 & 124.5${}^{b}$ \\ 
$\omega_5 (A^{\prime \prime})$  & 3260.1 & 0.3& 3176.7 & 0.0 & 3121.3 & 0.0 &2968.3 \\ 
$\omega_6 (A^{\prime \prime})$ & 864.6 & 5.6 & 921.6 & 6.3 & 924.9 & 6.1&904 \\ \hline
 \hline
    \end{tabular}
 \begin{tablenotes}
    \item ($\omega_1$) \ce{C-H} stretching; ($\omega_2$) \ce{C-O} stretching; ($\omega_3$) \ce{CH_2} bending; ($\omega_4$) out-of-plane bending; ($\omega_5$) \ce{C-H} stretching; ($\omega_6$) \ce{CH_2} rocking.
     \item{a)} From ref.~\citenum{Job1969,Clouthier1983}. Frequencies without a decimal point are deduced from combination bands and uncorrected for anharmonicity.
      \item{b)} The out-of-plane bending mode has a shallow double-well potential which is not well represented by the harmonic model used here.
 \end{tablenotes}
\end{threeparttable}
\end{table}

EOM-CCSD* considerably improves on EOM-CCSD for the \ce{C-O} bond length, however for other parameters which exhibit smaller changes between EOM-CCSD and EOM-CCSDT, the EOM-CCSD* results do not consistently move in the correct direction. Trends with respect to experimental data outside of the \ce{C-O} bond are also not clear. Because the \ce{C-O} elongation is the most significant change, though, the progression structure for \ce{C-O} stretching in the vibronic spectrum should be better represented. When examining the computed adiabatic excitation energies, a notable decrease in excitation energy arises when comparing the energy calculated by EOM-CCSD to EOM-CCSD* and EOM-CCSDT. The incorporation of triple excitations in the wave function markedly decreases the adiabatic excitation energies. While the reported experimental $T_e$ value lies between the EOM-CCSD and EOM-CCSD* results, the significantly improved agreement with full EOM-CCSDT indicates a better treatment of electron correlation and geometric relaxation, while additional factors such as basis set completeness would be needed to more closely match experiment.

Harmonic vibrational frequencies and infrared intensities 
have been obtained by numerical differentiation of analytic 
energy derivatives and dipole moments at the optimized geometries and are presented in Table~\ref{tab:formaldehyde-frequency}. From these results, we can confirm that the optimized geometry corresponds to a minimum. The frequencies computed using EOM-CCSD* align closely with those from EOM-CCSDT, displaying a consistent trend converging towards experimental data.

\begin{table}[ht]
    \centering
\begin{threeparttable}
    \begin{tabular}{cS[table-align-text-post = false]S[table-align-text-post = false]S[table-align-text-post = false]S[table-align-text-post = false]S[table-align-text-post = false]}
\hline
\hline
Parameter  & \multicolumn{1}{c}{$r_\text{CH}$ (\AA)} &  \multicolumn{1}{c}{$r_\text{CN}$ (\AA)} & \multicolumn{1}{c}{$r_\text{NN}$ (\AA)} & \multicolumn{1}{c}{$\angle_\text{HCN}$ (${}^\circ$)} & \multicolumn{1}{c}{$T_e$ (eV)} \\
\hline
Ground state & 1.078 & 1.336 &  1.323& 116.6 &\\
EOM-CCSD   &    1.077  &    1.328   &   1.312  & 119.4 &2.513 \\
EOM-CCSD* &      1.077   &   1.329   & 1.319  &   119.2& 1.760  \\
CASSCF${}^{a}$   & 1.067 & 1.329 & 1.305 & 121.4& \\
CASPT2${}^{a}$  & 1.073 & 1.333 & 1.321 & 121.5& \\
Mk-MRCCSD${}^{b}$ &      & 1.325 & 1.309 & &2.642 \\
Expt.${}^{c}$  & 1.063 & 1.324 & 1.349 & 123.2&  2.136${}^{d}$\\
Expt.${}^{e}$ & & 1.358(10) & 1.28(2) & 118.5(14) & \\
\hline
\hline
    \end{tabular}
    \caption{Optimized geometry of $s$-tetrazine $1{}^1\text{B}_\text{3u}$ ($n \xrightarrow{} \pi^*$) state with the basis set cc-pVTZ. Experimental uncertainties in the last significant digit(s) are given in parentheses.}
    \label{tab:AG-tetrazine}
 \begin{tablenotes}
     \item {a)} Ref.~\citenum{Schtz1995}
     \item {b)} Ref.~\citenum{Jagau2012}
     \item {c)} Ref.~\citenum{Innes1988}
     \item {d)} In solid benzene, Ref.~\citenum{Meyling1974}
     \item {e)} Ref.~\citenum{Smalley1977}
 \end{tablenotes}
\end{threeparttable}
\end{table}

\subsection{$n \xrightarrow{} \pi^*$ states of $s$-tetrazine}

Excited states of $s$-tetrazine are of great interest in the scientific community, as neatly summarized in the introduction of the work by Angeli\cite{Angeli2009}. 
The ground state geometry was obtained from the QUEST2 dataset\cite{Loos2019}, which has been optimized at the CC3/aug-cc-pVTZ level. 
In this study, we focus on the two lowest excited singlet states of $\text{B}_\text{3u}$ and $\text{B}_\text{2g}$ symmetry.
The $\text{B}_\text{3u}$ $n \xrightarrow{} \pi^*$ excited state is predominantly characterized by the configuration where a single electron is promoted from the highest occupied molecular orbital $b_{3g}$ to the lowest unoccupied molecular orbital $a_u$. 
Previous research has demonstrated that the first excited state exhibits $\text{D}_\text{2h}$ symmetry \cite{Schtz1995}, thus guiding our optimization search to only symmetric structures. 
The optimized geometries and adiabatic excitation energies are succinctly presented in Table~\ref{tab:AG-tetrazine}. Consistent with previous studies\cite{Schtz1995,Stanton1996-st,Jagau2012}, the change in geometry upon electronic excitation is small.
While minimal disparities are noted in the optimized geometries between the EOM-CCSD and EOM-CCSD* calculations, there is a small but consistent elongation in the aromatic bond lengths when going from EOM-CCSD to EOM-CCSD*, in opposition to the general contraction of all bonds relative to the ground state. As previously observed, the inclusion of triple excitations substantially reduces the adiabatic excitation energy by 0.75 eV. For comparison, results from additional theoretical studies and experimental findings are included in Table~\ref{tab:AG-tetrazine}.

The $\text{B}_\text{2g}$ $n \xrightarrow{} \pi^*$ excited state is characterized by a mixed configuration of single excitation $4b_{2u} \xrightarrow{} 1a_{u}$ and double excitation $1b_{1g}3b_{3g} \xrightarrow{} 1a_{u}1a_{u}$\cite{Angeli2009}. The optimized geometries are summarised in Table~\ref{tab:AG-tetrazine-mixed}. In contrast to the $\text{B}_\text{3u}$ excited state where small contractions of all bonds upon excitation are predicted, the $\text{B}_\text{2g}$ excited state relaxes to a geometry with a $\geq 0.13$ \AA{} elongation of the \ce{N-N} bond length and a $\geq 0.015$ \AA{} contraction of the \ce{C-N} bond length. The elongation of the \ce{N-N} bond can be partially explained by the excitation from bonding $1b_{1g}$ $\pi$ orbital to the anti-bonding $1a_{u}$ $\pi$ orbital. Similar to previous observations, EOM-CCSD* results in larger aromatic bond lengths compared to EOM-CCSD. The triples correction reduces the adiabatic excitation energy by 0.917 eV. This larger drop in energy upon introduction of triples likely reflects the increased double-excitation character in the excited state.

\begin{table}[!hb]
    \centering
    \begin{tabular}{cccccc}
\hline
\hline
Parameter  &   $r_\text{CH}$ (\AA) &  $r_\text{CN}$ (\AA) & $r_\text{NN}$ (\AA) & $\angle_\text{HCN}$ (${}^\circ$) &$T_e$ (eV)\\
Ground state & 1.078 & 1.336 &  1.323& 116.6 \\
EOM-CCSD   &    1.081   &    1.319   &   1.453    &    117.2  &5.162 \\
EOM-CCSD* &      1.081   &   1.321    &   1.456 &   117.1 &4.245  \\
\hline
\hline
    \end{tabular}
    \caption{Optimized geometry of $s$-tetrazine $1{}^1\text{B}_\text{2g}$ ($n \xrightarrow{} \pi^*$) state mixed with double excitation with basis set cc-pVTZ. }
    \label{tab:AG-tetrazine-mixed}
\end{table}

\subsection{$1{}^{1}\text{A}^{\prime}$ state of cytosine}
\begin{figure}
    \centering
    \includegraphics[width=0.9\textwidth]{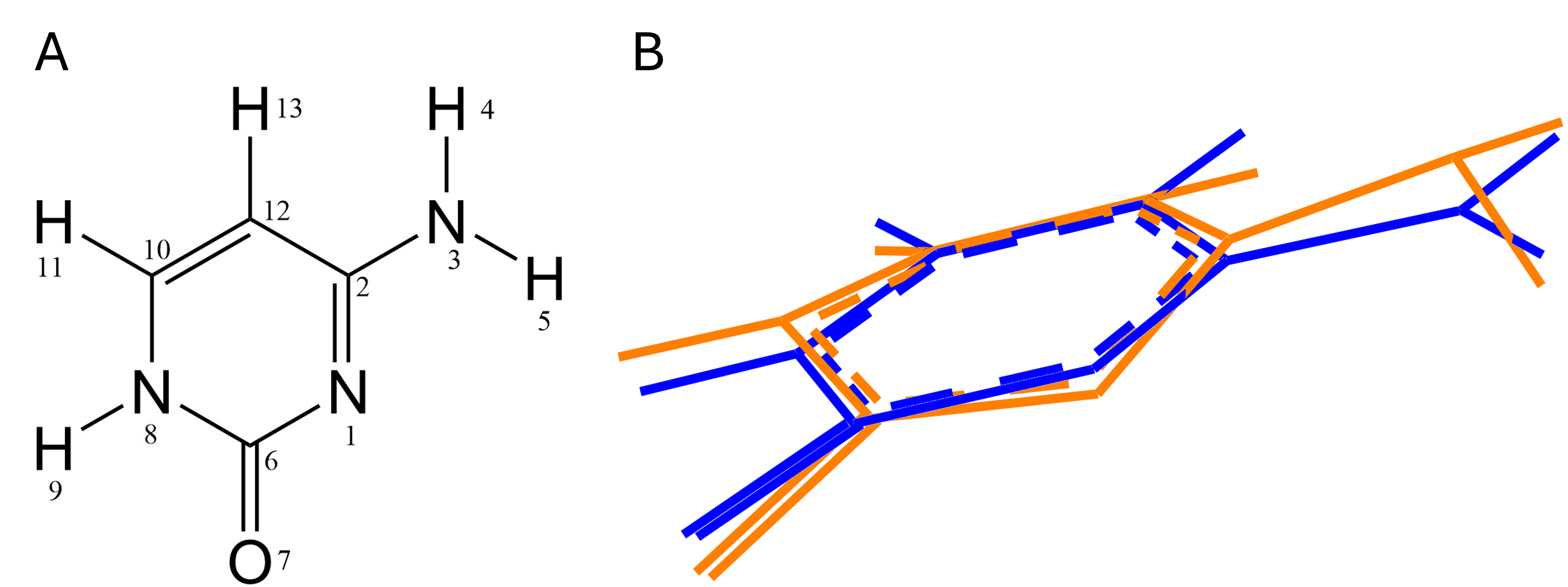}
    \caption{Cytosine structures. (A) Cytosine structure labeled with atom indices. (B) The geometric comparison between the ground state (blue) and optimized excited state with EOM-CCSD* (orange), molecules are in the same orientation as shown in A.}
    \label{fig:cytosine-structure}
\end{figure}

Cytosine, as one of the building blocks of life, has evoked keen interest in multiple areas\cite{CrespoHernndez2004}. It has been the motivation of experimental and theoretical efforts to characterize the nature and the properties of the lowest electronically excited states\cite{Shukla1999,Blancafort2004,CrespoHernndez2004,SerranoAndrs2005,Tajti2009,Szalay2012,Triandafillou2013,YaghoubiJouybari2020}. The lowest bright singly-excited state, which is of $\pi \xrightarrow{} \pi^*$ character, is of interest in this work. It is recognized that the optimization of excited states frequently leads to relaxed non-planar geometry\cite{CrespoHernndez2004,Lobsiger2013}. Thus $\text{C}_1$ symmetry is used during optimization. Cytosine presents a major challenge to numerical differentiation in terms of computational cost, as it has 33 internal degrees of freedom (and typically the numerical gradient for all 36 non-translational degrees of is calculated). Here, the ground state geometry is optimized at the MP2/cc-pVTZ level, while the excited state is optimized with the smaller cc-pVDZ basis set. The optimized geometries are summarized in Table~\ref{tab:cytosine-opt}, and the molecular structure of cytosine is shown in Figure~\ref{fig:cytosine-structure}. 
For EOM-CCSD, only geometries with RMS force convergence below $10^{-5}$ a.u. are presented, and are constrained to a planar geometry. Continuous unconstrained optimization results in significant distortion of the ring structure, making convergence and accurate assignment challenging, particularly due to mixing with the lowest $n \xrightarrow{} \pi^*$ state. For EOM-CCSD*, geometries with RMS force convergence below $10^{-7}$ a.u. are presented. The more facile convergence of EOM-CCSD* could be due to the more accurate relative energies of the $\pi \xrightarrow{} \pi^*$ and $n \xrightarrow{} \pi^*$ states which sensitively controls the strength of the pseudo-Jahn-Teller (PJT) effect and hence the extent of electronic mixing\cite{Stanton1996}. 
EOM-CCSD and EOM-CCSD* results show consistent trends in bond length changes. While the \ce{C6-N1} and \ce{C6-N8} bonds contract, all other bonds exhibit elongation. Aligning with prior research\cite{Ismail2002,Blancafort2004,Triandafillou2013}, the decay of the $\pi \xrightarrow{} \pi^*$ state involves the carbonyl bond stretching and the pyramidalization of \ce{C10} and \ce{N8}. Further frequency calculation confirmed it as a minimum (see Table 3 in Supporting information). The largest differences in bond lengths between EOM-CCSD and EOM-CCSD* are on the order of 0.06 \AA, while bond angles differ by as much as $9.3^\circ$ (although the bond angle differences are likely overstated due to the restriction of the EOM-CCSD geometry to planar).

\begin{table}
    \centering
    \begin{tabular}{cSSS}
\hline
\hline
Parameter& \multicolumn{1}{c}{Ground state} & \multicolumn{1}{c}{EOM-CCSD} & \multicolumn{1}{c}{EOM-CCSD*} \\
\hline
$r_{\ce{C2-N1}}$                      &  1.314   & 1.417  & 1.440    \\ 
$r_{\ce{N3-C2}}$                      &  1.354   & 1.376  & 1.401    \\ 
$r_{\ce{H4-N3}}$                      &  1.000   & 1.008  & 1.020    \\ 
$r_{\ce{H5-N3}}$                      &  1.004   & 1.009  & 1.021    \\ 
$r_{\ce{C6-N1}}$                      &  1.373   & 1.323  & 1.289    \\ 
$r_{\ce{O7-C6}}$                      &  1.218   & 1.269  & 1.331    \\ 
$r_{\ce{N8-C6}}$                      &  1.414   & 1.410  & 1.383    \\ 
$r_{\ce{H9-N8}}$                      &  1.008   & 1.015  & 1.014    \\ 
$r_{\ce{C10-N8}}$                   &  1.351   & 1.384  & 1.411    \\ 
$r_{\ce{H11-C10}}$               &  1.080   & 1.089  & 1.091    \\ 
$r_{\ce{C12-C10}}$               &  1.355   & 1.436  & 1.420    \\ 
$r_{\ce{H13-C12}}$               &  1.077   & 1.094  & 1.096    \\ 
$\angle_{\ce{N1-C2-N3}}$                  &  117.0   & 112.2  & 112.5    \\ 
$\angle_{\ce{C2-N3-H4}}$                  &  121.8   & 122.0  & 112.7    \\ 
$\angle_{\ce{C2-N3-H5}}$                  &  118.0   & 117.8  & 110.1    \\ 
$\angle_{\ce{C2-N1-C6}}$                  &  120.0   & 116.1  & 114.5    \\ 
$\angle_{\ce{N1-C6-O7}}$                  &  125.2   & 123.5  & 121.9    \\ 
$\angle_{\ce{N1-C6-N8}}$                  &  116.1   & 123.6  & 125.8    \\ 
$\angle_{\ce{C6-N8-H9}}$                  &  115.0   & 115.5  & 117.0    \\ 
$\angle_{\ce{C6-N8-C10}}$                 &  123.8   & 121.8  & 118.4    \\ 
$\angle_{\ce{N8-C10-H11}}$             &  117.0   & 118.7  & 118.6    \\ 
$\angle_{\ce{N8-C10-C12}}$             &  119.8   & 115.9  & 115.2    \\ 
$\angle_{\ce{C10-C12-H13}}$        &  121.4   & 119.1  & 119.6    \\ 
$\phi_{\ce{N1-C2-N3-H4}}$              &  180.0   & 180.0  & 155.0    \\ 
$\phi_{\ce{N1-C2-N3-H5}}$              &  0.0     & 0.0    & 31.7    \\ 
$\phi_{\ce{C6-N1-C2-N3}}$              &  180.0  & 180.0  & 163.1    \\ 
$\phi_{\ce{C2-N1-C6-O7}}$              &  180.0   & 180.0 & 177.8    \\ 
$\phi_{\ce{C2-N1-C6-N7}}$              &  0.0     & 0.0    & 3.1    \\ 
$\phi_{\ce{N1-C6-N8-H9}}$              &  180.0   & 180.0 & 178.7    \\ 
$\phi_{\ce{N1-C6-N8-C10}}$           &  0.0     & 0.0    & 18.1    \\ 
$\phi_{\ce{C6-N8-C10-H11}}$       &  180.0   & 180.0  & 155.3    \\ 
$\phi_{\ce{C6-N8-C10-C12}}$       &  0.0     & 0.0    & -22.6    \\ 
$\phi_{\ce{N8-C10-C12-H13}}$    &  180.0   & 180.0	 & 171.5    \\
\hline
\hline
    \end{tabular}
    \caption{Optimized bond lengths ($r$, \AA), bond angles (\angle, ${}^\circ$), and dihedral angles ($\phi$, ${}^\circ$) of the cytosine $ \pi \xrightarrow{} \pi^*$ state with basis set cc-pVDZ. EOM-CCSD RMS forces are converged to $10^{-5}$ a.u., see SI for convergence details; EOM-CCSD* RMS forces are converged to $10^{-7}$ a.u.}
    \label{tab:cytosine-opt}
\end{table}

\section{Conclusion}

It has been shown that the effect of triple excitations is important to achieving $\sim 0.1$ eV accuracy of transition energies\cite{Szalay2012}.  Despite numerous variants that approximately integrate triples effects into EOM-CCSD which have emerged over the years, the absence of analytic gradients remains a challenge. In response, this study introduces the formulation and implementation of analytic nuclear gradients for EOMEE-CCSD*. 

We showcase the applicability and efficiency of this advancement through demonstrative applications on formaldehyde, $s$-tetrazine, and cytosine systems. 

By streamlining the development and implementation of analytic gradient theory for this method, our work not only facilitates in-depth investigations with better accuracy but also substantially broadens the horizons for formulating analytic gradients and consistent properties for complex electronic structure methods spanning both the ground and excited states.

\begin{acknowledgement}

This work was supported by the US Department of Energy under grant DOE-SC0022893, and in part by the US National Science Foundation under grant CHE-2143725. All calculations were performed on the ManeFrame III computing system at SMU. 

\end{acknowledgement}

\section*{Supporting Information}
The following electronic supplementary information files are available from the publisher's website:
\begin{itemize}
\item SI.pdf: Supplementary information tables for the formaldehyde and cytosine calculations.
\item geometries.xlsx: Relevant geometries from this work in Cartesian coordinates.
\end{itemize}

\bibliography{eom}

\end{document}